\documentclass{aa}
\usepackage{graphicx}

\begin{document}

\hyphenation{be-san-con mo-dele }

\title{Early re-brightening of the afterglow of GRB~050525a
\thanks{Based on observations performed with
TAROT at the Calern observatory
}
}

\author{A. Klotz\inst{1,2}
        \and M. Bo\"er\inst{2}
        \and J.L. Atteia\inst{3}
        \and G. Stratta\inst{2,3}
        \and R. Behrend\inst{4}
        \and F. Malacrino\inst{3}
        \and Y. Darmedji\inst{1,2}
}

\institute{
CESR, Observatoire Midi-Pyr\'en\'ees (CNRS-UPS), BP 4346, F--31028 -
Toulouse Cedex 04, France
\and Observatoire de Haute Provence,
    F--04870 Saint Michel l'Observatoire, France
\and LAT, Observatoire Midi-Pyr\'en\'ees (CNRS-UPS), 14 Avenue E. Belin,
F--31400 - Toulouse, France
\and Observatoire de Gen\`eve, CH--1290 Sauverny, Switzerland
}

\offprints{A. Klotz, \email{klotz@cesr.fr}}

\date{Received {\today} /Accepted }

\titlerunning{Early re-brightening of the afterglow of GRB~050525a}
\authorrunning{Klotz {\it et al.}}

\abstract{
We present time resolved optical data acquired
by the TAROT automated observatory on the afterglow of GRB~050525a
from 6 to 136 minutes after the GRB. We evidence a rapid re-brightening
of 0.65 magnitude of the afterglow at $\sim$ 33 min after the GRB.
The decay slope $\alpha$ is $1.14\pm 0.07$ in the first part and is $1.23\pm 0.27$
after the re-brightening event. The afterglow of GRB~050525a is the third
known afterglow that exhibits a re-brightening event begining at 0.01--0.02 day
in the rest time frame.
}
 \maketitle

\keywords{gamma-ray : bursts }
\section{Introduction}

GRB~050525a was a very bright gamma-ray burst (GRB) detected on the 25$^{th}$ of May 2005
at 00:02:53.3 UT (hereafter $t_\mathrm{trig}$) by the BAT instrument on the Swift spacecraft
(trigger=130088, Band et al. \cite{band2005}). The gamma-ray light
curve
shows that GRB~050525a is a multipeak GRB,
with an emission lasting approximately 10 sec above 50 keV.
The fluence of GRB~050525a in the range 20-1000 keV is 7.84~10$^{-5}$ erg/cm$^2$,
and its peak energy is $E_{\rm peak}$ = 84 keV (Golenetskii et al. \cite{golenetskii2005}).
Spectroscopic observations performed 11 hours after the GRB~revealed
absorption lines from the host galaxy at a redshift z=0.606 (Foley et al. \cite{foley2005}).
At that redshift, and adopting a flat cosmology with $\Omega_m = 0.3$,
$\Omega_{\Lambda} = 0.7$, and h$_0 = 0.65$,
the isotropic-equivalent energy of GRB 050525a
in the range 1 keV to 10 MeV is $E_{\rm iso} = 12.6$ $10^{52}$ erg.
The intrinsic peak energy of the time integrated spectrum is 135$\pm$2 keV.

In this letter we report the early optical observations
of the GRB~050525a afterglow,
performed with the robotic TAROT observatory.
The Gamma-ray bursts Coordinates Network (GCN) notice, providing celestial
coordinates to ground stations, was send at 00:08:48 (Band et al. \cite{band2005}),
too late to detect the hypothetical optical prompt emission.
The first image of TAROT started at 00:08:52.1 UT, 5min59s after the
GRB. The afterglow was detected on all images taken until the end
of the night at the TAROT observatory (02:19 UT) at the coordinate quoted
by Rykoff et al.\cite{Rykoff2005}: R.A. 18h 32m 32.76s and Dec. +26$^{\circ}20'22.65''$ (J2000.0).
These data provide a continuous follow-up from 6 to 136 minutes
after the GRB. In this paper we show that the classical exponential decay
was perturbated at $t-t_\mathrm{trig} \simeq$ 33 min by a re-brightening event.

Section \ref{obsdata} describes the technical details
of the TAROT observations and of data reductions.
In section \ref{comparison}, we compare our early time observations
of GRB~050525a with those of other bursts with early optical observations
(t $\sim$ 0.01-0.1 day) and dense sampling.
In section \ref{discussion} we discuss
the theoretical interpretations which
have been proposed to explain the early re-brightening of GRB~optical afterglows.

\section{TAROT observations}
\label{obsdata}

TAROT is a fully autonomous 25 cm aperture
telescope installed at the Calern observatory (Observatoire de la
Cote d'Azur - France).
This telescope is
devoted to very early observations of GRB~optical counterparts.
A technical description of TAROT can be read in Bringer et
al. (\cite{Bringer99}) and in Bringer et al. (\cite{bringer2001}). The
CCD camera is a commercial Andor based on a Marconi 4240 chip and is placed at the
newtonian focus.
The spatial sampling is 3.3 arcsec/pixel.
The field of view is $1.86\degr$.
The readout noise is 9 electrons rms.
The readout time is 5 seconds
(to read the entire 2048x2048 matrix with no binning).

The first image was taken less than 4 seconds after the
position of GRB~050525a was provided by the GCN. A series of 76
images of various exposure times (15,
30, 60, 120\,s) were performed without any filter (hereafter clear filter). On-line
preprocessing software enable to provide calibrated images
less than 2 minutes after they were taken
(corrected by dark, flat and astrometrically calibrated
from the USNO-A1.0 catalog).

\begin{figure}[htb]
\includegraphics[width=\columnwidth]{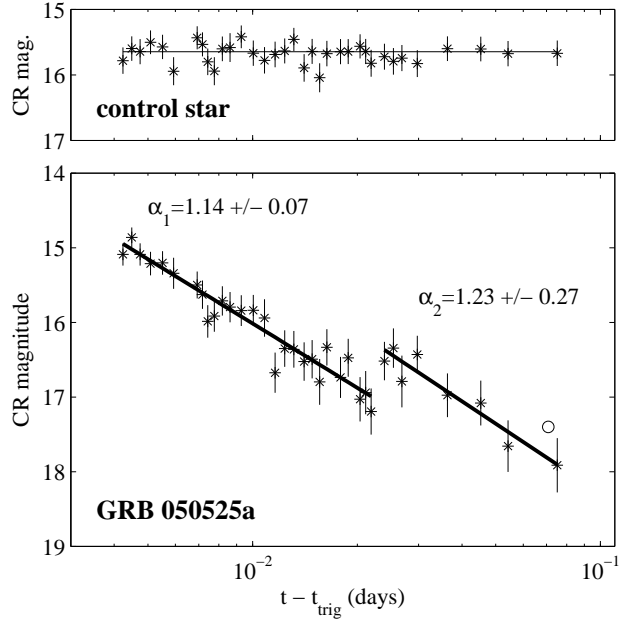}
\caption{
Top: Photometry of a field star (CR$\sim$15.65) assumed to be constant.
Bottom: GRB~050525a afterglow light curve obtained by TAROT
(data from Table \ref{logobstable}).
Straight lines are linear fits in the logarithm time scale
and their corresponding slope values. Error bars are 2$\sigma$
uncertainties (95\% of confidence).
The symbol $\circ$ is a R measurement
performed by Malesani et al. (\cite{malesani2005}) with the
Italian 3.6m TNG and calibrated with nearby USNO stars.
} \label{lc}
\end{figure}

On the first 31 images, the afterglow is bright enough
to be measured with a good accuracy on individual images.
Later images were co-added to increase the signal to noise.
Due to the decreasing of flux during exposures, the mean date
of an observation, $T$,
is not the middle of exposure; it must be interpolated between $t_1$
(start of the first frame) and $t_2$ (end of the last frame)
such that the flux $f$ verify:
\begin{eqnarray}
\int_{t_1}^{t_2} f(t)\,\mathrm{d}t = f(T)\,\int_{t_1}^{t_2} \mathrm{d}t
\label{eq1}
\end{eqnarray}
\\
Considering afterglow decay flux law: $f(t) \propto t ^{-\alpha}$, and
assuming all times counted
since $t_\mathrm{trig}$,
\begin{eqnarray}
T=\frac{t_2-t_1}{\ln(t_2/t_1)} & &\mbox{if $\alpha$=+1} \\
T=\left[\frac{(t_2-t_1)\cdot(1-\alpha)}{t_2{}^{1-\alpha}-t_1{}^{1-\alpha}}\right]^{1/\alpha}
& &\mbox{if $\alpha\neq$+1}
\label{eq2}
\end{eqnarray}
\\
$\alpha$ and $T$ values are computed by iterations.
Initial $T$ values are computed taking $\alpha$=+1. Then the fit
of the first light curve refines the $\alpha$ value. The second iteration
is enough to converge (see Table \ref{logobstable}
and Fig.~\ref{lc}).

\begin{table}[htb]
\caption{ Log of the measurements. The first column is the $T$ date
from GRB~(in minutes) as defined in formula \ref{eq1}.
The second is the CR magnitude and the third
is the error.}
\begin{center}
\begin{tabular}{c c c | c c c}
$T$ (min) & CR      & 2$\sigma$ & $T$ (min) & CR      & 2$\sigma$ \cr
\noalign{\smallskip} \hline \noalign{\smallskip}
 6.104  & 15.09   & 0.15    & 18.931  & 16.36   & 0.25    \cr
 6.471  & 14.86   & 0.13    & 20.249  & 16.52   & 0.26    \cr
 6.839  & 15.09   & 0.15    & 21.351  & 16.49   & 0.26    \cr
 7.326  & 15.21   & 0.16    & 22.452  & 16.80   & 0.31    \cr
 7.931  & 15.20   & 0.16    & 23.554  & 16.33   & 0.24    \cr
 8.536  & 15.34   & 0.21    & 25.779  & 16.74   & 0.27    \cr
 9.978  & 15.50   & 0.18    & 27.111  & 16.47   & 0.25    \cr
10.338  & 15.63   & 0.19    & 29.351  & 17.03   & 0.30    \cr
10.698  & 15.99   & 0.22    & 30.453  & 16.94   & 0.29    \cr
11.172  & 15.91   & 0.21    & 31.569  & 17.19   & 0.31    \cr
11.791  & 15.71   & 0.20    & 34.477  & 16.52   & 0.26    \cr
12.403  & 15.80   & 0.20    & 36.580  & 16.34   & 0.26    \cr
13.363  & 15.84   & 0.21    & 38.697  & 16.79   & 0.35    \cr
14.472  & 15.84   & 0.21    & 42.842  & 16.43   & 0.25    \cr
15.589  & 15.94   & 0.25    & 52.338  & 16.98   & 0.29    \cr
16.705  & 16.67   & 0.27    & 65.275  & 17.08   & 0.30    \cr
17.814  & 16.35   & 0.25    & 78.185  & 17.66   & 0.34    \cr
18.931  & 16.36   & 0.25    & 108.318  & 17.91   & 0.36    \cr
\noalign{\smallskip} \hline
\end{tabular}
\label{logobstable}
\end{center}
\end{table}

The presence of the star
USNO-B1 1163-0325216 (R$\sim$16.6), at about
15 arcsec north-west from the afterglow, perturbated the
photometry, especially when the afterglow fades.
The set of magnitudes provided by Klotz et al. (\cite{klotz2005})
was affected by this effect, which leads to a false
plateau for dates 40 min after the trigger. To eliminate
this effect, we computed an image (designated hereafter {\it mask}) of the field
which does not show the afterglow. The {\it mask}
is synthetized in two steps: first
we oversampled the 76 images by a factor three and
stack them to synthetize the sum. In
this image, the afterglow and the star are well separated.
The second step is to clear only the afterglow spot in order to synthetize
the {\it mask}. The {\it mask} was normalized in flux for each
image and was substracted, leading to images where only
the afterglow appears (and some residuals of bright
stars). Then we can extract the magnitude of
the afterglow avoiding problems of contamination
by the nearby star.

The GRB~030329 afterglow shows there is no colour
effect during the first phases of decay (Zeh et al. \cite{zeh2003a}, \cite{zeh2003b}).
Taking this advantage, we performed differential photometry
with two stars as reference: USNO-B1 1163-0325130 (R=11.25) and
USNO-B1 1163-0325158 (R=14.00).
First, we verified that the magnitude of reference stars
does not vary.
As we have no information about the afterglow colours (we used no filter),
we do not obtain directly R magnitudes. We designated by CR, the
unfiltered magnitudes calibrated by USNO-B1 R magnitudes of reference stars.
As the airmass only varied from 1.13 to 1.05
during measurements, its effect on colours
is assumed to be negligible (presumatly lower than 0.05 mag, much less
than other uncertainties).
As a consequence, it is possible to convert our CR
magnitudes into standard R magnitudes by a simple
offset (estimated lower than 0.2 magnitude, depending on the
intrinsic colour differences between afterglow and reference stars).
Anyway, the CR magnitudes allows the compute decay
slope parameter $\alpha$.


Light curve (Fig. \ref{lc}) shows two parts
separated at $t-t_\mathrm{trig}\simeq$ 33 min. Fits
of slopes give $\alpha_1=1.14\pm 0.07$ and
$\alpha_2=1.23\pm 0.27$. Uncertainties are such
that slopes are not significantly differents.
The most important remark is the offset of about 0.65
magnitude (nearly a factor two) between the two parts. We can join
the two curves by two extreme paths: i) a sharp re-brightening of 0.65
magnitude in less than 3 minute centered at
$t-t_\mathrm{trig}=$ 32.8 min, ii) a plateau (flat re-brightening)
of CR~$\simeq$~16.7, begining at
$t-t_\mathrm{trig}\simeq$ 26 min and finishing at $t-t_\mathrm{trig}\simeq$ 43 min.
The magnitude uncertainties of our measurements are
too large to discriminate one assumption to the other.

On Fig. \ref{lc}, we reported a measurement ($\circ$) obtained by
Malesani et al. (\cite{malesani2005}). They used
a large aperture telescope and a R filter. They
found the afterglow 0.5 magnitude brighter than
CR TAROT value. Less than 0.2 mag. of this offset
can be due to the CR-R differences as explained
in section~\ref{obsdata}. The remaining 0.3 mag
may be due to the substraction of the
USNO-B1 1163-0325216 nearby star.
Due to a much larger aperture of telescope,
Malesani mesurement is probably
best accurate than the two last measurements
obtained with TAROT. As a consequence
$\alpha_2$=1.23 is probably over estimated.

To summarize, the re-brightening by a factor two of the afterglow of GRB~050525a
is effective, the transition duration is comprised from less
than 3 minutes ($\delta t /t = 0.1$) up to 17 minutes ($\delta t /t = 0.5$)
centered at $t-t_\mathrm{trig}=$ 32.8 min, and the slope of the
temporal decay before and after the re-brightening are fully comparable.

\begin{figure}[htb]
\includegraphics[width=\columnwidth]{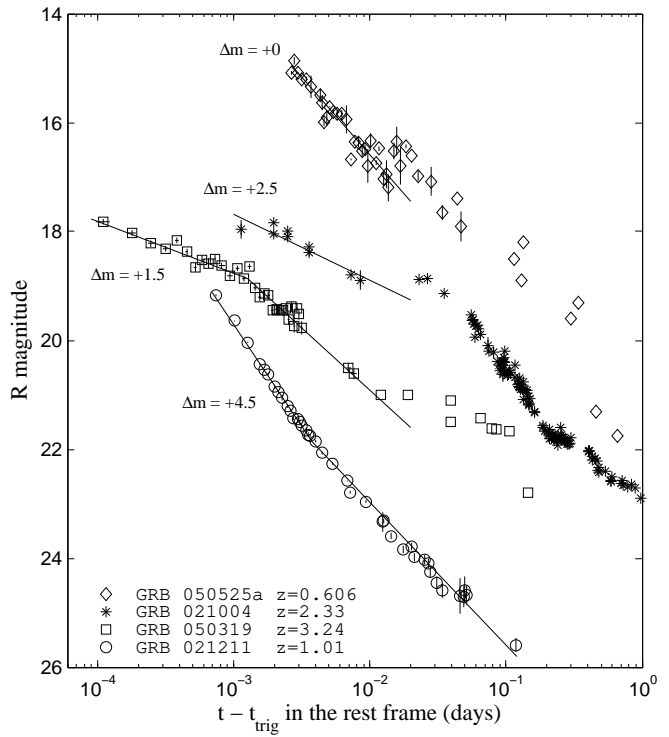}
\caption{
Optical light-curves of some early afterglows scaled in their rest time frame.
Curves are vertically shifted by $\Delta$m values to avoid crossing of curves.
The upper three curves show re-brightening events at $\sim$ 1--2~10$^{-2}$ day.
Data for GRB~050525a are from TAROT, and GCNCs
3465, 3469, 3470, 3489, 3491, 3488, 3493, 3486, 3506.
Data for GRB~021004 are from
Fox et al.~(\cite{fox2003}) and GCNCs
1564, 1566, 1570, 1573, 1576, 1577,
1578, 1580, 1581, 1582, 1584, 1587, 1591, 1594, 1606, 1614, 1615, 1628.
Data for GRB~050319 are from
Wozniak et al.~(\cite{wozniak2005}) and GCNCs
3120, 3121, 3124, 3130, 3139.
Data for GRB~021211 are from
Holland et al.~(\cite{holland2002}),
Li et al.~(\cite{li2003}),
Fox et al.~(\cite{fox2003b}).
}
\label{afterglows}
\end{figure}

\section{Early optical GRB afterglows}
\label{comparison}

Only a few afterglows have been observed at optical wavelengths
less than one hour after the GRB.
They include GRB 990123, GRB 020418, GRB 021004 (z=2.33), GRB 021211 (z=1.01),
GRB 041219a, GRB 050319 (z=3.24), GRB 050502a (z=3.79), and GRB 050525a (z=0.606)
discussed in this paper.
Within this small sample, short-scale variability seems to be the
rule rather than the exception.
This variability can be observed as single or multiple re-brightenings
(GRB~021004, GRB 041219a, GRB 050319, GRB 050525a),
as a shallowing (GRB~990123, GRB~021211)
or a steepening(GRB~050319) of the light curve, or as a
gradual rise of the afterglow (GRB~030418).

Figure~\ref{afterglows} compares the early optical afterglows
of four GRBs with a measured redshift (GRB 050502a is not included
in this comparison since the data on its early afterglow have not yet been published).
For a proper comparison, the abscissa of the plot gives the time after the trigger,
{\it in the referential of the GRB}. A striking feature is the presence
of an episode of re-brightening starting 0.01 to 0.02 days after the trigger, in three out
of the four GRBs displayed in figure~\ref{afterglows}.\footnote{GRB 041219a,
whose redshift is not known, exhibits a re-brightening
episode starting $\sim$17 min, or 0.012 day, after the trigger (Blake et al. 2005).}
At first glance the three GRBs which exhibit rebrightnening
episodes do not have special properties. GRB 040319 is a single pulse
GRB, while GRB 021004 and GRB 050525a are multi-peak events.
The isotropic-equivalent energies and rest frame peak energies of GRBs
in figure \ref{afterglows} are 12.6 10$^{52}$ erg and 135 keV for GRB 050525a,
$\sim 3.1$ $10^{52}$ erg for GRB 050319\footnote{The peak energy of GRB 050319
is not known.}, 5.1 10$^{52}$ erg and 266 keV for GRB 021004, and
1.4 10$^{52}$ erg, and 92 keV for GRB 021211.
Since three out of four GRBs with early optical follow-up exhibit
re-brightening episodes, it is tempting to conclude that they
represent a common feature of GRB afterglows. This remark
clearly points out the necessity of very quick optical follow-up
to measure the decay slope before 0.01 day (14 min), required
to assess the time and amplitude of possible re-brightenings in
future GRBs.

\section{Discussion and conclusion}
\label{discussion}

Within the context of the internal/external shock model of GRBs, re-brightening episodes
have been explained by the reverse shock, by a continuing activity
of the central engine, by a variable density profile
of the external environment in which the fireball expands, by the presence
of neutrons in the ejecta, or by the destruction of dust surrounding the source.

Late afterglow emission is explained as the forward shock component of external shock
produced by the interaction of the expanding fireball with the external medium.
Very early optical emission is thought to be an effect of the reverse shock component of the external shock.
The emergence of the forward shock from the reverse shock emission component
produces a shallowing of the early-time light curve after an initial steep decay
that can mimick a re-brightening (Panaitescu \& Kumar \cite{panaitescu2004}).
Indeed this model has been recently invoked to explain the phenomenology of GRB 050525a (Shao \& Dai \cite{Shao05}).

Variable energy input model also predicts early light curve re-brightenings (Nakar et al. \cite{nakar2003}).
The energy variability could be provided by refreshed shocks produced by massive and slow
shells ejected late in the GRB,
that collide with the inital blast wave when it has decelerated.
After the collision of each shell, the flux from the fireball increases
but the light curve decay will continue with the same rate as prior to the collision.
The net result is a {\it shift} upward of the initial light curve at the time
of the collision with same decay rate as before collision (Bjornsson et al. \cite{bjornsson2004}).
Alternatively, a variable energy input coud be due to initial energy
(per solid angle) inhomogeneities in the jet ({\it patchy shell model} described by
Kumar \& Piran \cite{kumar2000b}).

In the variable density scenario, clumpy inter-stellar matter (ISM)
or variable wind expelled from the massive progenitor,
produces high density regions that,
interacting with the expanding fireball, could provide flux enhancements
if particular electron cooling conditions are satisifed
(e.g. Lazzati et al. \cite{lazzati2002}).
In this case, in fact, the flux would be sensitive to density variations only
in the 'slow cooling regime', with the observed frequency below the
cooling frequency (Sari et al. \cite{Sari98}). The recovery of the initial slope after the bump could be
achieved requiring a decrease in the density below the initial value immediately
after the high density region (Lazzati et al. \cite{lazzati2002}).

Afterglow re-brightening has been predicted also by the
neutron-fed afterglow model when the already decelerating ion shell
sweeps up the trail of decay products left from the ahead decaying
neutron shell. The arrival time of the re-brightening depends on the
Lorentz factor of the neutron shell and can vary from few seconds to
several tens of days after the burst (Beloborodov~\cite{Beloborodov2003}).

Finally, dust destruction mechanisms might provide an enhancement of the
observed optical flux within a time scale that depends, among several other
parameters, on the density distribution of the circumburst dust.
A decreasing of reddening is predicted in this case (Perna et al. \cite{perna2003}).

GRB 050525a is, after GRB 050319, the second GRB with a known distance,
with detailed observations of the early afterglow at optical (this paper)
and X-ray wavelengths (Band et al.~\cite{band2005}). This proves that, eigth years after the discovery
of GRB afterglows, a new window opens: multi-wavelength observations of
the very early afterglow (the first hour).
This remarkable advance has been possible thanks to the availability
of arcminute localizations quickly distributed to efficient robotic telescopes, and
to the excellent performance of the XRT on-board SWIFT.
The observations presented in this paper demonstrate the richness
of the information contained in the very early afterglow, and
the great promises of the multi-wavelength
observations which are now within our range.

\begin{acknowledgements}
The TAROT telescope has been funded by the {\it Centre National
de la Recherche Scientifique} (CNRS), {\it Institut National des
Sciences de l'Univers} (INSU) and the Carlsberg Fundation. It has
been built with the support of the {\it Division Technique} of
INSU. We thank the technical staff
joined with the TAROT project: G. Bucholtz, J. Esseric,
C. Pollas and Y. Richaud.
\end{acknowledgements}

\end{document}